\documentclass[twoside,english,nofootinbib,preprintnumbers,aps,onecolumn,notitlepage,11pt]{revtex4}
\usepackage{euscript,amssymb}
\usepackage{amsthm}
\usepackage{graphicx}
\usepackage{amsfonts}
\usepackage{amsmath}
\usepackage{amssymb}
\usepackage{fancyhdr}
\usepackage{epsfig}
\usepackage{color}
\usepackage{soul}
\usepackage{enumitem}
\usepackage{calc}
\usepackage{enumerate}
\usepackage{afterpage}
\usepackage{multirow}
\usepackage[papersize={8.5in,11in}]{geometry}

\allowdisplaybreaks

\newcommand{\be}{\begin{equation}}
\newcommand{\ee}{\end{equation}}

\begin{document}

\title{Petrov type D equation on horizons of nontrivial bundle topology}
\author{Denis Dobkowski-Ry{\l}ko}
	\email{Denis.Dobkowski-Rylko@fuw.edu.pl}
	\affiliation{Faculty of Physics, University of Warsaw, ul. Pasteura 5, 02-093 Warsaw, Poland}

\author{Jerzy Lewandowski}
	\email{Jerzy.Lewandowski@fuw.edu.pl}
	\affiliation{Faculty of Physics, University of Warsaw, ul. Pasteura 5, 02-093 Warsaw, Poland}
\author{Istv\'an R\'acz}
	\email{Istvan.Racz@fuw.edu.pl}
	\affiliation{Faculty of Physics, University of Warsaw, ul. Pasteura 5, 02-093 Warsaw, Poland}
	\affiliation{Wigner RCP, H-1121 Budapest, Konkoly Thege Mikl\'{o}s \'{u}t 29-33, Hungary}
\begin{abstract} 
     We  consider  $3$-dimensional isolated horizons (IHs) generated by null curves  that form  nontrivial $U(1)$ bundles. We find a natural interplay between the IH geometry and the $U(1)$-bundle geometry. In this context we consider the Petrov type D equation introduced and studied in  previous works \cite{DLP1,DLP2,LS,DKLS1}. From the $4$-dimensional spacetime point of view, solutions to that equation define isolated horizons embeddable in vacuum spacetimes (with cosmological constant) as Killing horizons to the second order such that the spacetime Weyl tensor at the  horizon is of the Petrov type D. From the point of view of the $U(1)$-bundle structure, the equation couples a $U(1)$-connection, a metric tensor defined on the  base manifold and the surface gravity in a very nontrivial way. We focus  on the $U(1)$-bundles over $2$-dimensional manifolds diffeomorphic to $2$-sphere. We have derived all the axisymmetric solutions to the Petrov type D equation. For a fixed value of the cosmological constant they set a $3$-dimensional family as one could expect. A surprising result is, that generically our horizons are not embeddable in the known exact solutions to Einstein's equations. It means that among the exact type D spacetimes there exists a new family of spacetimes that generalize the properties of the Kerr- (anti) de Sitter  black holes on one hand and the Taub-NUT spacetimes on the other hand.
\end{abstract}

\date{\today}

\pacs{???}

\maketitle
\section{Introduction} 
The theory of  non expanding horizons (NEH) is often used to describe black holes \cite{ABL1}. It is, however, far more general and may also be applied to spacetimes containing cosmological horizons, null boundaries of the conformally completed asymptotically flat spacetimes \cite{AM} or black hole holograph construction of spacetimes about isolated horizons \cite{R1, R2, LRS}. Properties of NEHs find their analogs in the black hole spacetimes, such as the black hole 'thermodynamics' \cite{ABL2}, uniqueness theorems \cite{LP2} and rigidity theorem \cite{LP1}.  The long term program is to understand conditions satisfied by geometry of NEHs that distinguish the horizons of physical black holes. In a case of NEH embeddable in spacetime as a Killing horizon to the second order, the vacuum Einstein equations (possibly with a cosmological constant) and the Petrov type D of the spacetime Weyl tensor at the horizon,  amount to an equation on the Riemann geometry  induced on the 2-dimensional space of null generators and the 2-form representing the rotation \cite{LP3,DLP1}. The equation and solutions were investigated   in the case of horizons  which have the structure of a trivial principal fiber bundle.  For rotating solutions,   the only allowed topology of a  cross-section is that of a $2$-sphere \cite{DKLS1}.  For bifurcated horizons,  the type D equation implies the axial  symmetry \cite{raczKillingSpinor,LS}.  All the axially symmetric solutions were derived \cite{LP3,DLP2}. For every value of the cosmological constant, they form a $2$-dimensional family that can be parametrized by the area and angular momentum.   In that sense,  the equation has the properties of rigidity and no hair so well known in the global black hole theory \cite{HE}.  Eventhough the Petrov type D equation was derived for non-extremal horizons, it is also an integrability condition for the condition  satisfied by  the geometry and rotation $1$-form potential  induced on  $2$-dimensional spaces of null generators of extremal Killing horizons to the first  order  \cite{DLP1,ABL1,LP2}. This condition is also known as the Near Horizon Geometry (NHG) equation \cite{Kunduri1,Kunduri2}. That relation between the type D  and the NHG equation was used to show that in the case of non-zero genus,  the only solutions to the NHG equation  are geometries of constant Gauss curvature and zero rotation $1$-form potential \cite{DKLS2}.                        

In this paper we consider the Petrov type D equation and the vacuum Einstein equations with cosmological constant for  IHs of the structure of a nontrivial bundle, the Hopf bundle or, more generally, the Dirac monopol bundle. Hence,  the  space of the null generators is  topologically $2$-sphere, however there is no global spacelike cross-section. An example of spacetime containing such horizon is the Taub-NUT spacetime \cite{PG}. We derive all the axisymmetric solutions to the Petrov type D equation. They set a $3$-dimensional family for every value of the cosmological constant. As it could be expected, there  emerges  a new parameter - the topological charge times the surface gravity. The final result, however, is surprising. In the previous, trivial bundle case, the axisymmetric Petrov type D horizons  are embeddable in the Kerr-(anti) de Sitter spacetimes. The generic horizons we find  in the current case turn out  not to be embeddable in the known in the literature generalizations of the Kerr spacetimes. That conclusion is so surprising, because the Kerr-NUT spacetimes contain horizons of the Petrov type D that are believed  to have the Hopf (or Dirac monopol) bundle structure.          

\section{Isolated horizons  of nontrivial $U(1)$-bundle  topology}
In this section  we introduce a general definition of $3$-dimensional isolated horizons (IHs) whose null generators have  the structure of nontrivial fibration. While eventually the horizons are surfaces  in $4$-dimensional spacetimes, their intrinsic structure can be considered on its own, independently of an embedding. That is  what we do in the first subsection below.  In the second subsection, we explain the $4$-dimensional spacetime context of embedded IHs, the  symmetries assumed in this paper, briefly discuss the Einstein constraints and recall   the Petrov type equation.  For a detailed derivation  of the Petrov type D equation for IHs see \cite{DLP1}. The derivation is  local  and  applies also to the current case.  

In this paper we use the same abstract index notation \cite{Wald}, as in \cite{DLP1}:
\begin{itemize}
\item  Indices of $4$-dimensional spacetime $M$ tensors  are denoted by lower Greek letters: $ \alpha, \beta,\gamma, ... \ = 1,2,3,4$.
\item  Tensors defined in $3$-dimensional  space $H$  carry indexes denoted by lower Latin letters: $a, b, c, ...\ = 1,2,3$.
\item  Capital Latin letters $A, B, C, ... \ = 1,2$ are for the tensors defined on the 2-dimensional space $S$ of the null generators of $H$.  
\end{itemize}

\subsection{IH structure on a  $U(1)$ bundle}       
A nontrivial bundle structure  is a new element introduced  in the IH theory in the current paper.  Let    
\be\label{HS} \Pi: H \rightarrow S\ee 
be a principal fiber bundle with the structure group $U(1)$.  Denote by $\ell$ the fundamental vector field  on $H$, that is such that  its flow coincides with the action of $U(1)$ on $H$.  We normalize $\ell$ such that the parameter of the flow ranges the interval  $[0, 2\pi]$. 

Throughout this paper 
\be {\rm dim}H=3 .\ee  

On  $H$ we introduce an IH geometry compatible with the bundle structure. It consists of: 

$(i)$ a degenerate metric tensor  $g_{ab}$ of the signature $0++$, such that 
\be \label{g}  \ell^ag_{ab} = 0 = {\cal L}_\ell g_{ab} ;  \ee    
and 

$(ii)$  a covariant derivative $\nabla_a$ on  $T(H)$, torsion free and  such that
\be \label{inv} \nabla_ag_{bc} =0, \ \ \ \ \ [{\cal L}_\ell,\nabla_a]=0 . \ee     
The second condition means that $\nabla_a$ is invariant with respect to the the action of the $U(1)$ group on $H$. The same is true about $g_{ab}$ due to the second equality in (\ref{g}) .   

It follows, that 
\be \ell^a\nabla_a \ell^b = \kappa\ell^b \ee
and through out this paper we are assuming that
\be\label{kappa} \kappa = {\rm const}\not=0.  \ee
After assuming the Einstein constraints, the constancy of $\kappa$ will be a necessary property, and the not-vanishing means that $H$ is a non-extremal (non-degenerate) IH.    

The key ingredient of the covariant derivative (for our paper) is the rotation $1$-form potential $\omega_a$ defined as follows
\be  \label{omega} \nabla_a\ell^b = \omega_a\ell^b ,  \ee
and by (\ref{inv}) it satisfies
\be \label{Llomega}  {\cal L}_\ell{\omega}_a=0.\ee 
It follows from (\ref{kappa}) that, the $1$-form
\begin{align}\label{tomega}
\tilde{\omega}:= \frac{1}{\kappa}\omega 
\end{align}
is a connection $1$-form on the $U(1)$  bundle (\ref{HS}). Indeed, due to (\ref{omega}) and the second eq. (\ref{inv})    $\omega_a$ satisfies
\be \ell^a\tilde{\omega}_a=1, \ \ \ \ \ \ \ {\cal L}_\ell \tilde{\omega}_a=0 .  \ee 
The degenerate metric tensor $g_{ab}$ induces on $S$ a (genuine) metric tensor $g_{AB}$ such that $g_{ab}$ is its pullback,
\be g_{ab} = \Pi^*{}_{ab}{}^{AB}g_{AB}.  \ee

The area $2$-form $\eta_{AB}$ defined on $S$ and corresponding to $g_{AB}$ (and some orientation of $S$) may also be pulled back to $H$,
\be  \eta_{ab} := \Pi^*{}_{ab}{}^{AB}\eta_{AB} .\ee
We use it to define a rotation pseudo scalar $\Omega$, 
\begin{align}\label{Omega1}
\Omega  \eta_{ab} := d\omega_{ab} = \kappa d\tilde{\omega}_{ab}.
\end{align}
It satisfies 
\be  \ell^a \Omega_{,a} = 0\ee
hence we  consider $\Omega$   as a function on $S$.    The rotation $1$-form potential $\omega_a$ can be represented by, locally defined in a neighborhood of each point $x\in S$, $1$-forms $\omega_A$ such that 
\be  d\omega_{AB} = \Omega \eta_{AB},  \ee   
where $\Omega$ is a scalar function globally defined and regular on the entire manifold $S$.
\subsection{Embedded IHs and the  Petrov type D equation}
In $4$-dimensional spacetime  $(M,g_{\mu \nu})$ of the signature $-+++$, every IH  $(H,\ell^a,g_{ab}, \nabla_a)$ introduced in the previous subsection is a null surface such that the intrinsic geometry $(g_{ab},\nabla_a)$ coincides with   the spacetime metric tensor $g_{\mu\nu}$ and the covariant derivative $\nabla_\mu$   restricted to $T(H)$ (preserved by $\nabla_\mu$). Due to the intrinsic symmetries  (the second eq. (\ref{g}) and the second eq. (\ref{inv}))   IH $H$ can be called a Killing horizon to the first order. Indeed, there exists an extension $t$ of the vector field  $\ell$ to a neighborhood of $H$ in $M$, such that 
 \be\label{Kill}  {\cal L}_tg_{\mu\nu}{}_{|_H}=0=[{\cal L}_t, \nabla_\mu]{}_{|_H} . \ee
   
Throughout this paper we assume that the spacetime metric tensor $g_{\mu \nu}$ satisfies the vacuum Einstein equations:
\begin{equation}
G_{\mu \nu} +\Lambda g_{\mu\nu} = 0,
\end{equation}
where $\Lambda$ is a cosmological constant and $G_{\mu \nu}$ is the Einstein tensor. 

The constraints induced on  $(H,\ell^a,g_{ab}, \nabla_a)$ by  Einstein's   equations are soluble explicitly   in the non-extremal case (\ref{kappa}) considered in this paper. The  degenerate metric tensor $g_{ab}$ and the rotation $1$-form $\omega_a$  can be set freely on $H$ (modulo (\ref{g},\ref{kappa},\ref{Llomega})), and they determine the remaining ingredients of  $\nabla_a$.  

Henceforth, about  the vector field $t$ (\ref{Kill}) and the spacetime Weyl tensor $C_{\mu\nu\alpha\beta}$ we assume a stronger 
condition, namely
\be \label{Kill2}  {\cal L}_tC_{\mu\nu\alpha\beta}{}_{|_H}=0. \ee
That property of $H$ may be called a Killing (or stationary isolated) horizon to the second order. That assumption does not constraint the intrinsic non-extremal IH $H$ geometry $(g_{ab}, \nabla_a)$. Instead, via the Einstein equations, it determines all the components of the spacetime Weyl tensor  at $H$  by $g_{ab}$ and $\omega_a$.  Therefore assumption that the Petrov type of the Weyl tensor at $H$ is D turns into an equation on $(g_{AB}, \Omega)$. We consider that equation in the next section.  However before doing that  we would like to briefly give the idea on the spacetime elements of the problem and  sketch the derivation  presented in \cite{DLP1}.  

In the IH framework we use  adapted  null frames (see \cite{DLP1}, Sec. III.B ). Each of them consists of a real null vector field $\ell^\mu$ tangent to $H$ and coinciding thereon with the vector field $t^\mu$ (\ref{Kill}), another real null vector field $n^\mu$ orthogonal to a foliation of $H$ by space like 3-sections, and two complex valued vector fields $m^\mu$ tangent to the foliation.  The foliation and the frame are preserved by the flow generated by the vector field $\ell^\mu$.   The Weyl tensor complex components (see \cite{DLP1}, Sec. III.C )  $\Psi_0, \ \Psi_1, \ \Psi_2$ and $\Psi_3$ are automatically guaranteed to be constant along the null generators of $H$ (with the first two vanishing). So the condition (\ref{Kill2})  is on $\Psi_4$ only.  Moreover, in the consequence of the Bianchi identities the components $\Psi_3$ and now also $\Psi_4$ can be expressed by $\Psi_2$ and its first and second derivatives. Next, the Weyl tensor is of the Petrov type D iff it admits two  double  principal null directions (PNDs). The vector field $\ell^\mu$ is already tangent to a double PND (the vanishing of $\Psi_0$ and $\Psi_1$). Hence, the type D assumption concerns the remaining transversal PNDs. For a generic type D horizon,  the transversal null vector $n^\mu$ is not a PND for any adapted null frame. However, the existence of a second double PND turns into an algebraic condition, namely
\be   3\Psi_2\Psi_4 - 2\Psi_3{}^2=0 .  \ee 
 After expressing $\Psi_3$ and $\Psi_4$ by  $\Psi_2$ and its derivatives, the equation somewhat magically takes a compact form. We  remind it in the next section. 

\section{The Petrov type D equation}
Given an IH structure  $(H, \ell^a, g_{ab}, \nabla_a, S)$, introduced above, the Petrov type D equation is imposed on the Riemaniann metric $g_{AB}$ and  the rotation pseudo-scalar  $\Omega$ defined on $S$. We will also use  the Gauss  curvature $K$ of $g_{AB}$,
\begin{align}\label{Omega}
{}^{(2)}R_{AB} =: K g_{AB}, 
\end{align}
where ${}^{(2)}R_{AB}$ is  the Ricci tensor  of the metric tensor $g_{AB}$. 
To write the equation we introduce a complex null co-frame $m_A$ such that the metric $g_{AB}$ and area 2-form $\eta_{AB}$ take the following form:
\begin{align}
g_{AB} = m_A \bar m_B + m_B \bar m_A, && \eta_{AB} = i(\bar m_A m_B - \bar m_B m_A).
\end{align}
The Weyl tensor is of the type D along the generator $\Pi^{-1}(x)$ of the horizon $H$,  if and only if the following equation, which we refer to as the type D equation, holds true at the point $x\in S$:
\begin{equation}\label{typeDgen}
\bar m^A \bar m^B {}^{(2)}\nabla_A {}^{(2)}\nabla_B \left( K - \frac{\Lambda}{3} + i\Omega \right)^{-\frac{1}{3}} = 0,
\end{equation}
where ${}^{(2)}\nabla_A$ is the torsion free, metric covariant derivative defined by $g_{AB}$, 
and the term in the bracket does not vanish at the point $x$, namely:
\begin{align}
K-\frac{\Lambda}{3}+i\Omega \ne 0.
\end{align}
This function is related to the only non-zero invariant (given $\ell$) component 
\begin{align}\label{psi2}
\Psi_2 = -\frac{1}{2}\left(K+i\Omega \right)+ \frac{\Lambda}{6}.
\end{align}
of the type D Weyl tensor, and if that component vanishes, then all the Weyl tensor vanishes at that point.  

We will solve eq. (\ref{typeDgen}) assuming that the base manifold $S$ (\ref{HS}) is diffeomorphic to 2-sphere,
\be S=S_2.\ee In that case, all the $U(1)$ bundles  are numbered by integers. An  integer $m$ corresponding to $H$ can be calculated from the curvature of the $U(1)$-connection  $1$-form $\tilde{\omega}$, that passes to a condition on the rotation pseudo-invariant $\Omega$
\begin{align}\label{intOmega} 
 \int_{S_2}\Omega \eta_{AB}= 2\pi  m\kappa =: 2\pi n.
\end{align}
For each $\Omega$ there exist 1-forms $\omega^+$ and $\omega^-$ defined on $S_2$ apart from the  southern and northern pole respectively such that:
\begin{align}\label{slomega}
d\omega^{\pm}_{AB} = \Omega\eta_{AB}.
\end{align}
Incidentally, from the mathematical point of view, the case $\kappa=1$ seems to be the most interesting. However we do not see any reason implied by GR, that would restrict $\kappa$ to that value.  

We also assume, that the metric tensor  $g_{AB}$ and the rotation pseudo-scalar $\Omega$ invariantly defined on $S$ admit an axial symmetry.  Consequently, we choose the coordinates adapted to the symmetry (Appendix) in which the 2-metric tensor $g_{AB}$ reads:
\begin{align}\label{gAB}
g_{AB}dx^A dx^B = R^2\left( \frac{1}{P(x)^2} dx^2+ P(x)^2d\varphi^2 \right),
\end{align}
where $x \in [-1;1]$, $\varphi \in [0; 2\pi]$ and $R$ is the area parameter \cite{LP2, AEPB}. The frame vector and its dual take the form:
\begin{align}
m^A\partial_A=\frac{1}{R\sqrt{2}}\left( P(x)\partial_x + i \frac{1}{P(x)} \partial_\varphi \right), && \bar m_A dx^A = \frac{R}{\sqrt{2}}\left( \frac{1}{P(x)}dx - iP(x)d\varphi \right).
\end{align}
The above coordinate system is not well defined at $x=\pm1$, therefore to derive the regularity conditions\footnote{for more details see Appendix.} we use the relation between the metric (\ref{gAB}) and the standard 2-sphere coordinates, namely:
\begin{align}
R^2\left( \frac{1}{P(x)^2} dx^2+ P(x)^2d\varphi^2  \right) = \Sigma^2(\theta) \left( d\theta^2 + \sin^2\theta d\varphi^2  \right).
\end{align}
For an axisymmetric scalar function $f$ defined globally on $S_2$ (as one of the functions: $K$, $\Omega$ and $\Psi_2$) the differentiability condition at the poles $x=\pm 1$ reads:
\begin{align}
\partial_\theta f= 0.
\end{align}
This condition is equivalent to:
\begin{align}
P\partial_xf=0
\end{align}
and it will be assumed for functions $K$ and $\Omega$ (or in other words for $\Psi_2$). For the metric (\ref{gAB}) to be twice differentiable at the poles the following boundary conditions must be satisfied:
\begin{align}\label{cond1}
P^2\big |_{x= \pm 1} &= 0, \\ \label{cond2}
 \partial_x P^2 |_{x= \pm 1} &= \mp 2. 
 \end{align}
The condition (\ref{intOmega}) boils down to
 \begin{align}\label{cond3}
\int^1_{-1} \Omega R^2 dx &=  m \kappa =  n.
\end{align}
Eq. (\ref{typeDgen}) in the coordinates adapted to the axial symmetry reads
\begin{align}
\partial_x^2\Psi_2=0,
\end{align}
and its general solution is of the form:
\begin{align}\label{Psi2}
\Psi_2 = (c_1 x+ c_2)^{-\frac{1}{3}},
\end{align}
where $c_1$ and $c_2$ are complex constants. Now comparing it to eq. (\ref{psi2}) and expressing the Gaussian curvature in the introduced coordinates yields:
\begin{equation}\label{typeD}
\frac{1}{(c_1 x+c_2)^3} = \frac{1}{4R^2}\partial^2_xP^2 - \frac{1}{2}i\Omega +\Lambda'.
\end{equation}
This equation can be solved for the values of  the complex parameters  $c_1, c_2$ that satisfy solubility conditions.

\section{Solution to the Petrov type D equation on the nontrivial bundle topology}
In the following section we will solve eq. (\ref{typeD}), first for the case in which the constant $c_1$ vanishes, and later for $c_1$ taking arbitrary (nonzero) complex values. In the case when $c_2$ vanishes the geometry is not well defined (see below) therefore we will exclude it from our considerations. We have used a similar approach in \cite{DLP2}, where we solved the type D equation for the trivial bundle that is $n=0$.

\subsection{The solution for vanishing $c_1$}
The type D equation (\ref{typeD}) with the vanishing complex constant $c_1$ reads:
\begin{equation}\label{typeD0}
\frac{4R^2}{c_2^3} = \partial^2_xP^2 - 2iR^2 \Omega +4R^2\Lambda'.
\end{equation}
where $\Lambda'$ (as in \cite{DLP2}) denotes a rescaled cosmological constant: 
\begin{equation}
\Lambda' := \Lambda/6. 
\end{equation}
Integrating both sides of eq. (\ref{typeD0}) and using boundary conditions (\ref{cond2}) and (\ref{cond3}) yields:
\begin{equation}
c_2^3=\frac{4R^2}{-2-in+4\Lambda' R^2}.
\end{equation}
We then find the solution to eq. (\ref{typeD0}):
\begin{equation}\label{vanP^2}
P^2 = 1-x^2,
\end{equation}
and
\begin{equation}
\Omega=\frac{n}{2R^2}.
\end{equation}
Now we find the rotation 1-form potential $\omega^\pm$. Since $\omega^\pm$ has to satisfy eq. (\ref{slomega}) and the regularity conditions at $x=\pm1$, namely:
\begin{align}\label{omegacond}
\omega^+|_{x=1} = 0 = \omega^-|_{x=-1},
\end{align}  
it follows that
\begin{align}
\omega^\pm_A dx^A = \frac{n}{2} (x \mp 1)d\varphi.
\end{align}
Consequently, in case of $c_1=0$, solution to the type D equation can be parametrized by 2 parameters: $R^2$ and $n$. Moreover, if $n=0$ then $\omega^\pm$ vanishes. The found solution is embeddable in the Taub-NUT-(anti) de Sitter spacetime which is of the type D and is defined by the static spacetime metric tensors satisfying the vacuum Einstein equations with the cosmological constant $\Lambda$ \cite{PG}, namely:
\begin{align}\label{tnds}
ds^2 = -\frac{Q}{\rho^2}\left[ dt -4l\sin^2\left(\tfrac{1}{2}\theta \right)d\phi \right]^2 + \frac{\rho^2}{Q}dr^2+ \rho^2\left( d\theta^2 +\sin^2\theta d\phi^2 \right),
\end{align}
where
\begin{align}
\rho^2 &= r^2+ l^2; \\
Q &= r^2-2Mr -l^2 - \Lambda\left( -l^4+2l^2r^2 + \tfrac{1}{3}r^4 \right).
\end{align}
Its extension contains Killing horizons, which are parametrized by the roots of the equation:
\begin{align}
r_H^2-2Mr_H -l^2 - \Lambda\left( -l^4+2l^2r_H^2 + \tfrac{1}{3}r_H^4 \right) = 0.
\end{align}
Each of such horizons, that is not non-extremal,  is one of the type D horizons that we consider. The 2-metric on the (space of the null generators of) Killing horizon admits spherical symmetry:
\begin{align}
ds_2^2=\rho^2\left( d\theta^2 +\sin^2\theta d\phi^2 \right)
\end{align}
and the relation between cordinates $x$, $\varphi$ and $\theta$, $\phi$ is the following:
\begin{align}
x(\theta)=-\cos\theta, && \varphi(\phi) = \phi.
\end{align}
Furthermore, we express the parameters $R^2$ and $n$ in terms of the parameters of the Taub-NUT horizon, namely $r$ and $l$. From the comparison of the area of the $S_2$ metrics on the horizon, we find that:
\begin{align}\label{Rrl}
R^2 = r_H^2 + l^2.
\end{align} 
The Killing vector field:
\begin{equation}
\xi=M \frac{\partial}{\partial t},
\end{equation}
on the horizon defines our generator of the null symmetry (we have introduced the factor $M$ to make the vector field dimensionless as above).
It is such that on the horizon:
\begin{align}
 \ell = \xi_{|H}.
\end{align}
Next we use the following formula
\begin{align}
{(\xi^\mu \xi_\mu)_{;\nu}}_{|H}= -2\kappa \xi_\nu
\end{align}
to calculate the surface gravity $\kappa$ on the horizon:
\begin{align}\label{kappanut}
\kappa &= \frac{-M}{2r_H}\left(\frac{-\Lambda r_H^4+(1-2\Lambda l^2)r_H^2+(1-\Lambda l^2)l^2}{r_H^2+l^2}\right).
\end{align}
In case of the Taub-NUT-(anti) de Sitter metric (\ref{tnds}), the 1-form $\tilde{\omega}^-$ reads:
\begin{align}\label{tildeomeganut}
\tilde{\omega}^-_A dx^A = -\frac{4l}{M}\sin^2\left(\tfrac{1}{2}\theta \right)d\phi.
\end{align}
Furthermore, we use the obtained expressions for $\kappa$ (\ref{kappanut}) and $\tilde\omega$ (\ref{tildeomeganut}) and plug them into (\ref{tomega}), (\ref{intOmega}) and (\ref{slomega}) to find the relation between $n$ and the parameters $r$ and $l$, namely
\begin{align}\label{nlr}
n = \frac{-4 l \kappa}{M} =  \frac{2l}{r_H}\left(\frac{-\Lambda r_H^4+(1-2\Lambda l^2)r_H^2+(1-\Lambda l^2)l^2}{r_H^2+l^2}\right)
\end{align}
To conclude, the found horizon for the vanishing $c_1$ is embeddable in the quotient of Taub-NUT-(anti) de Sitter spacetime by the symmetry: $t \mapsto t+ 2\pi M$ and the correspondence between our parameters and those of the Taub-NUT-(anti) de Sitter horizon is listed in (\ref{Rrl}) and (\ref{nlr}). The embedding, obviously, is not unique and depends on the chosen symmetry.

\subsection{The solution for arbitrary nonzero $c_1$}
Now assuming that neither complex constant vanishes we integrate eq. (\ref{typeD}) twice to obtain:
\begin{equation}\label{P^2}
P^2=2R^2 Re\left[ \frac{1}{c_1^2(c_1 x+ c_2)}\right] - 2R^2 \Lambda' x^2 + Cx + D.
\end{equation}
Using the boundary conditions (\ref{cond1}) and (\ref{cond2}) we find integration constants $C$ and $D$:
\begin{align}\label{C}
C&= -2+4R^2\Lambda' + 2R^2 Re\left[ \frac{1}{c_1 (c_1+c_2)^2} \right]=-2R^2Re\left[\frac{1}{c_1(c_1^2-c_2^2)}\right], \\
D&=2R^2 Re \left[ \frac{2c_1+c_2}{c_1^2(c_1+c_2)^2} \right] +2R^2 \Lambda' - 2,
\end{align}
Moreover, integrating eq. (\ref{typeD}) once and using (\ref{cond2}) and (\ref{cond3}) yields the relation between $R^2$, $\Lambda'$, $n$ and parameters $c_1$, $c_2$:
\begin{equation}\label{R^2}
R^2 = \frac{-2-in}{4(\frac{c_2}{(c_1^2-c_2^2)^2}-\Lambda')}.
\end{equation}
The area radius $R^2$ has to be real (and positive), therefore the following equation must be satisfied:
\begin{equation}
Im\left[ \frac{4\left( \frac{c_2}{(c_1^2-c_2^2)^2} -\Lambda' \right)}{2+in} \right] = 0.
\end{equation}
Consequently, we can choose the following parametrization:
\begin{equation}
\frac{c_2}{(c_1^2-c_2^2)^2} = \frac{1}{\gamma}- i\frac{1}{2} n \left(\Lambda' - \frac{1}{\gamma} \right),
\end{equation}
where $\gamma$ is a real parameter. The last equality in (\ref{C}) yields:
\begin{equation}
\frac{1}{2R^2} - \Lambda' = Re \left[ \left( \frac{c_1}{c_2} -1 \right) \frac{c_2}{(c_1^2-c^2_2)^2} \right]
\end{equation}
which we use to introduce yet another real parameter $\eta$:
\begin{equation}
\frac{c_2}{c_1} = \frac{ \eta n}{4\Lambda' R^2-2} + i\eta = \frac{1}{2} \eta n (\Lambda' \gamma-1) + i \eta,
\end{equation}
where we have assumed that:
\begin{equation}
1-2\Lambda' R^2 \neq 0.
\end{equation}
Using such parametrization, the expression for $P^2$ reads:
\begin{equation}\label{P^2}
P^2 = \frac{\left( 1-x^2 \right)\left( \left( x-\frac{1}{2} \eta n \left( 1-\Lambda' \gamma \right) \right)^2 + \eta^2+ \frac{1-x^2}{1-\Lambda' \gamma} \right)}{\left( x-\frac{1}{2} \eta n \left( 1-\Lambda' \gamma \right)\right)^2 + \eta^2}.
\end{equation}
For $n=0$ eq. (\ref{P^2}) reduces to the case known from \cite{DLP1, DLP2}, namely:
\begin{equation}
P^2 = \frac{\left( 1-x^2 \right)\left( \eta^2 \left( 1-\Lambda' \gamma \right)-\Lambda' \gamma x^2 + 1 \right)}{\left( 1-\Lambda' \gamma \right) \left(x^2+ \eta^2 \right)}.
\end{equation}
We can now calculate the rotation 1-form potential $\omega^\pm$, just as we previously did for $c_1=0$. Taking the imaginary part of both sides of the type D equation (\ref{typeD}) yields:
\begin{align}
\Omega &= Im \left[\frac{-2}{c_1^3\left(x+\frac{c_2}{c_1} \right)^3}\right] \nonumber\\
&=Im \left[ \frac{2i\left(1-\eta^2\left(\frac{1}{2}n\left(\Lambda'\gamma-1\right)+i\right)^2\right)}{\eta \gamma \left( x+\frac{1}{2}\eta n \left(\Lambda'\gamma-1\right)+i\eta\right)^3}\right]
\end{align}
and therefore
\begin{equation}
\omega^\pm = Im \left[ \frac{i \left( 1- \eta^2\left( \frac{n}{2}\left(\Lambda'\gamma -1\right)+i\right)^2\right)^2}{2\eta \left( 1-\Lambda' \gamma \right) \left( x+\eta\left( \frac{n}{2}\left(\Lambda'\gamma -1 \right) + i\right)\right)^2} +iC^{\pm}\right]d\phi.
\end{equation}
Since $\omega^\pm$ satisfies the boundary conditions (\ref{omegacond}) it follows that:
\begin{align}
C^\pm = \frac{1}{2\eta(1-\gamma \Lambda')} \left[ 1- \eta^2  +\frac{n^2\eta^2}{4} (1- \gamma \Lambda')^2 \mp n \eta (1-\gamma \Lambda')  \right].
\end{align}
Taking all into consideration, the family of solutions (for $c_1\neq0$) to the type D equation (\ref{typeD}) can be expressed in terms of three real parameters: $\eta$, $\gamma$ and $n$.

In case of:
\begin{equation}
1-2\Lambda' R^2 = 0
\end{equation}
we introduce the following parametrization:
\begin{align}
\frac{c_2}{(c_1^2-c_2^2)^2} = -\frac{1}{2}i n\Lambda' && \frac{c_2}{c_1} = - \frac{n \Lambda'}{2 \alpha}.
\end{align}
It is easy to see that both expressions vanish for $n=0$, that is consistent with the result obtained in \cite{DLP2}, in which $R^2=\frac{1}{2\Lambda'}$ case has been excluded for the geometry to be well defined. The frame coefficient $P^2$ takes the following form:
\begin{equation}\label{Pspecial}
P^2 = 1-x^2,
\end{equation}
and 
\begin{equation}
\Omega= -\frac{2\alpha\left( 1-\left(\frac{n \Lambda'}{2 \alpha}\right)^2 \right)^2}{\left(x-\frac{n \Lambda' }{2 \alpha} \right)^3}.
\end{equation}
The 1-form $\omega^\pm$ reads:
\begin{align}
\omega^\pm = \left[\frac{\alpha\left( 1- \left(\frac{n \Lambda'}{2 \alpha}\right)^2 \right)^2}{2\Lambda' \left( x-\frac{n \Lambda'}{2 \alpha}\right)^2} + C^\pm\right]d\varphi,
\end{align}
where
\begin{equation}
C^\pm = -\frac{\alpha}{2\Lambda'} \left( 1\pm \frac{n\Lambda'}{2\alpha}  \right)^2.
\end{equation}

\section{Summary} We have considered $3$-dimensional IHs (isolated horizons) generated by null curves  that form  nontrivial $U(1)$ bundles. In the non-extremal IH case, the rotation $1$-form potential corresponds to a connection on the bundle times the surface gravity. Hence, there is a natural interplay between the IH geometry and the $U(1)$-bundle geometry. In this context we have considered the Petrov type D equation (\ref{typeDgen}) introduced and studied in  previous works (\cite{DLP1,DLP2,LS,DKLS1}). From the $4$-dimensional spacetime point of view, solutions of that equation define isolated horizons embeddable in vacuum spacetimes (with cosmological constant) as Killing horizons to the second order such that the spacetime Weyl tensor at the  horizon is of the Petrov type D. From the point of view of the $U(1)$-bundle structure, the equation couples a $U(1)$-connection, a metric tensor defined on the  base manifold and the surface gravity in a very nontrivial way. An example of known spacetime containing an  IH of the nontrivial bundle structure  is the Taub-NUT solution. The Killing horizon in that spacetime has the structure of the Hopf fibration of $S_3$ over $S_2$, and it is of  the Petrov type D (along with all the spacetime).  In the current paper we have focused on the $U(1)$-bundles over $2$-dim manifolds diffeomorphic to $2$-sphere.  A general bundle of that type is characterized by an integer topological charge and is mathematically equivalent to the Dirac monopol, however,  the role of the electromagnetic vector potential of the original Dirac monopole in our case is played by the rotation $1$-form potential divided  by the surface gravity.   We have derived all the axisymmetric solutions to the Petrov type D equation.   Below we summarize our results.  The  analysis is followed by our final comments. 

The solutions we have derived are  determined by the cosmological constant $\Lambda$, the area radius $R^2$,  a function $P$ ($R$ and $P$  give rise to the metric (\ref{gAB})),  and by  the rotation pseudo-scalar $\Omega$ (\ref{Omega1}).   From $g_{AB}$ and $\Omega$ one can reconstruct the remaining element of the IH, namely derivative $\nabla_a$ (see \cite{DLP1}).  The topological charge $m$ (integer) of the $U(1)$-bundle structure of the horizon and the surface gravity $\kappa$ set the parameter $n$ that features in Table I. The list of $(\Lambda, R^2, P, \Omega)$  we have found is divided into three classes. We discuss them now.  

The first class consists of  the metric tensors $g_{AB}$  of constant Gaussian curvature 
\be K=\frac{1}{R^2}\ee 
and constant rotation scalar $\Omega$ related in the table with $n$ and $R^2$, and is embeddable in the Taub-NUT-(anti) de Sitter spacetime. The cosmological constant is arbitrary in this class, and unrelated to $K$ and $\Omega$. Hence, that class is parametrized freely by three real parameters $\Lambda',R^2>0$ and $n$.

Class 2 in the Table I is characterized by the special relation between $R^2$ and $\Lambda =:6\Lambda'$,
\begin{align}
R^2=\frac{1}{2\Lambda'},
\end{align}
and by the condition
\be  \partial_A\Omega \not=0 . \ee
The class is parametrized by real parameters $\Lambda', n, \alpha$ constrained by certain conditions discussed now. 
 It follows that here we can only consider positive $\Lambda'$ for the area radius $R^2$ to be positive:
\begin{equation}
\Lambda'>0.
\end{equation}
 Furthermore, the frame coefficient takes the form (\ref{Pspecial}) and it is clear that it is non-negative for all $x$ in the domain. However, one has to pay attention to the behavior of $\Psi_2$, eq. (\ref{Psi2}), on the domain $x\in[-1,1]$ and require it to be well-defined, which means:
 \begin{align}
  \left|\frac{n\Lambda'}{2\alpha}\right|>1.
 \end{align}
The third class is  parametrized by real parameters $\Lambda',  \eta, \gamma, n$. 
First, we specify their domains, in which the metric $g_{AB}$ is  well-defined and at least $4$-times differentiable including the poles of the sphere. We want the area radius $R^2$ to be positive and therefore:
\begin{align}
R^2 = \frac{1}{2} \frac{\gamma}{\Lambda' \gamma - 1} > 0 && \Leftrightarrow && \Lambda'>\frac{1}{\gamma}.
\end{align}
Also the frame coefficient $P^2$ has to be positive for $x\in(-1,1)$:
\begin{align}
P^2>0 &\Leftrightarrow  \left( x-\frac{1}{2} \eta n \left( 1-\Lambda' \gamma \right) \right)^2 + \eta^2+ \frac{1-x^2}{1-\Lambda' \gamma}>0 
\end{align}
and that occures whenever one of the following is satisfied:
\begin{itemize}[leftmargin=*]
\item $\gamma < 0$;
\item $\Big(\gamma > 0\Big)  \wedge \bigg( \eta^2 > \frac{-\Lambda' \gamma}{\left(1-\Lambda' \gamma\right)\left( \left(1-\Lambda' \gamma \right)^2\frac{n^2}{4}+\Lambda' \gamma\right)}\bigg)$;
\item $\Big(\gamma > 0\Big)  \wedge \bigg( \eta^2 < \frac{-\Lambda' \gamma}{\left(1-\Lambda' \gamma\right)\left( \left(1-\Lambda' \gamma \right)^2\frac{n^2}{4}+\Lambda' \gamma\right)}\bigg)\\ \wedge \bigg(\big|\eta n \big|< \frac{\Lambda'\gamma+\sqrt{\left( (1-\Lambda'\gamma)\eta^2+1 \right)\Lambda'\gamma+\frac{1}{4}\eta^2n^2(1-\Lambda'\gamma)^3}}{\frac{1}{2}(1-\Lambda'\gamma)}\bigg)$.
\end{itemize}
Moreover, $c_2$ in (\ref{Psi2}) has to be nonzero, because otherwise $\Psi_2$ component of the Weyl tensor would be ill-defined at $x=0$, it follows that $\eta$ cannot vanish at any case:
\begin{equation}
\eta\neq0.
\end{equation} 

Several remarks are in order. 

The first remark concerns reconstruction of a $U(1)$-bundle and the IH structure from the data provided above. 
Let us fix arbitrarily a topological charge 
\be m\not= 0 \ee 
and a corresponding $U(1)$-bundle $\Pi:H\rightarrow S_2$.   Then, for every data from the table such that 
\be n\not=0 \ee
we set the surface gravity $\kappa$ to be
\be \kappa = \frac{n}{m} ,\ee
and we can reconstruct a unique (modulo automorphisms of $H$) IH structure $g_{ab}, \nabla_a$.  

For 
\be n=0 \ee
on the other hand,  Table I reduces to the earlier derived \cite{DLP2} axisymmetric  solutions to the Petrov type D equation the horizon of the $\mathbb{R}\times S^2$ topology.  Those horizons can be defined by a subgroup of $\mathbb{R}$ and become the trivial $U(1)$ bundle $U(1)\times S_2$. 

The last remark concerns the  issue of embedding  the generic IHs of the Petrov type D found in the current paper in the known exact solutions to Einstein's equations. Only the special class I family of solutions is embeddable in the known spacetime of the type D, namely  the Taub-NUT-(anti) de Sitter spacetime. In the trivial bundle case of $H= \mathbb{R}\times S_2$ considered in previous works \cite{DLP2}, a generic axisymmetric  Petrov type D IH  $(H, g_{ab}, \nabla_a)$  can be embedded  in one of the non-extremal  Schwarzschild - (anti) de Sitter / Kerr-(anti) de Sitter spacetimes. In the current case, however, for the nonzero values of $n$ and a generic solution in the Table I, we were not able to identify any generalized black hole solution that can accommodate it. That result  requires a better understanding. It may be an indication of an existence of a new family of Kerr -(anti) de Sitter - Taub-NUT like spacetimes.

\clearpage

\begin{center}
\begin{table}[]\label{table}
\begin{tabular}{|c|c|c|}
\hline
\multicolumn{3}{|c|}{Possible solutions to type D equation}                                                          \\ \hline
 {Class I} & {Class II} & {Class III}                \\ \hline \hline
{\multirow{2}{*}{\hspace{0.25cm}$R^2>0$}} & \multirow{2}{*}{\hspace{0.25cm}$R^2=\frac{1}{2\Lambda'}$ and $\Lambda'>0$} & \multirow{2}{*}{\hspace{0.25cm}$R^2\neq\frac{1}{2\Lambda'}$} \\
{\multirow{2}{*}{\hspace{3cm}}} & \multirow{2}{*}{\hspace{6cm}} & \multirow{2}{*}{\hspace{6cm}} \\
{\multirow{2}{*}{\hspace{0.25cm}$P^2=1-x^2$}} & \multirow{2}{*}{\hspace{0.25cm}$P^2=1-x^2$} & \multirow{2}{*}{\hspace{0.25cm}$P^2=\frac{\left( 1-x^2 \right)\left( \left( x-\frac{1}{2} \eta n \left( 1-\Lambda' \gamma \right) \right)^2 + \eta^2+ \frac{1-x^2}{1-\Lambda' \gamma} \right)}{\left( x-\frac{1}{2} \eta n \left( 1-\Lambda' \gamma \right)\right)^2 + \eta^2}$} \\
{\multirow{2}{*}{}} & \multirow{2}{*}{} & \multirow{2}{*}{} \\
{\multirow{2}{*}{\hspace{0.25cm}$\Omega=\frac{n}{2R^2}$}} & \multirow{2}{*}{\hspace{0.25cm}$\Omega= -\frac{2\alpha\left( 1-\left(\frac{n \Lambda'}{2 \alpha}\right)^2 \right)^2}{\left(x-\frac{n \Lambda' }{2 \alpha} \right)^3}$} & \multirow{2}{*}{\hspace{0.25cm}$\Omega = Im \left[ \frac{2i\left(1-\eta^2\left(\frac{1}{2}n\left(\Lambda'\gamma-1\right)+i\right)^2\right)}{\eta \gamma \left( x+\frac{1}{2}\eta n \left(\Lambda'\gamma-1\right)+i\eta\right)^3}\right]$} \\
{\multirow{2}{*}{}} & \multirow{2}{*}{} & \multirow{2}{*}{} \\
                  &                   &                   \\ \hline
\end{tabular}
\caption {Solutions to the type D equation on horizons of nontrivial bundle topology divided into three classes.}
\end{table}
\end{center}

\noindent{\bf Acknowledgements:}
This work was partially supported by the Polish National Science Centre grant No. 2017/27/B/ST2/02806.
IR and JL were supported by the POLONEZ programme of the National
Science Centre of Poland (under the project No. 2016/23/P/ST1/04195)
which has received funding from the European Union`s Horizon 2020
research and innovation programme under the Marie Sk{\l}odowska-Curie
grant agreement No.~665778.
\setlength\fboxrule{0pt}\setlength\fboxsep{0mm}{\fbox{
	\includegraphics[scale=0.07]{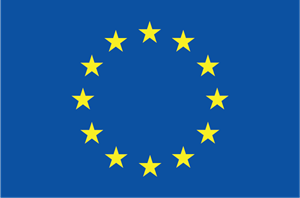}
} }

\section{Appendix}
The conditions (\ref{cond1}) and (\ref{cond2}) are necessary for the metric tensor (\ref{gAB}) to be continues and differentiable at  the poles \cite{DLP2}. Now, we will show (\ref{cond1}) and (\ref{cond2}) are also sufficient.
Consider a 2-sphere metric with a conformal factor $\Sigma$ (independent of $\varphi$ because of the symmetry):
 \begin{equation}\label{eq:2Smetric}
 g_{AB}dx^Adx^B= \Sigma^2(\theta) (d\theta^2 + \sin^2\theta d\varphi^2),
 \end{equation}
and the following transformation:
\begin{equation}
dx = \frac{\Sigma^2 \sin\theta}{R^2} d\theta,
\end{equation}
where $R^2$ is defined to be the area radius satisfying: 
$$A= 4\pi R^2.$$ 
We now introduce the frame coefficient: 
$$P^2=\frac{\Sigma^2 \sin^2{\theta}}{R^2}.$$ Calculating the area of the transformed metric $g_{AB}$ yields:
\begin{equation}
A = R^2 (x_1 - x_0) 2\pi.
\end{equation}
Since $x$ has been defined up to an additive constant, by setting $x_1=1$ from the above equation we obtain that $x_0=-1$. The coordinate $\varphi$ is such that the (normalized) infinitesimal axial symmetry equals $\partial_\varphi$ and the curves $\varphi=\text{const}$ are orthogonal to the infinitesimal symmetry. 
The metric tensor $g_{AB}$ reads: 
\begin{equation}\label{metric}
g_{AB}dx^Adx^B = R^2\bigg(\frac{1}{P(x)^2} dx^2 + P(x)^2 d\varphi ^2\bigg).
\end{equation}

Next, we check whether condition for the lack of conical singularity, namely:
\begin{align}
 \lim_{x\to \pm1} \partial_x P^2 = \mp 2
\end{align}
implies that the metric (\ref{eq:2Smetric}) is differentiable, that is if $\Sigma,_\theta=0$ is satisfied on the poles. Using the relation between $P$ and $\Sigma$ we obtain:
\begin{align}
\Sigma,_\theta &= \partial_\theta \left( \frac{PR}{\sin\theta} \right) = \frac{R P,_\theta - \Sigma \cos\theta}{\sin\theta} = \frac{\frac{\Sigma^2 \sin\theta}{R} P,_x - \Sigma \cos\theta}{\sin\theta} = \Sigma \frac{PP,_x - \cos\theta}{\sin\theta}
\end{align}
Now taking a limit as $\theta$ approaches $0$ (or $\pi$) and using the L'Hospital's rule we find:
\begin{align}\label{domega}
\Sigma,_\theta \big |_{\theta=0,\pi} = \lim_{\theta \to 0,\pi} \frac{RP(1+\frac{1}{2} \frac{P^2}{\sin^2\theta}\partial_x^2P^2)}{\cos\theta}.
\end{align}
As long as the limit of $\frac{P}{\sin\theta}$ as $\theta$ approaches $0$ (or $\pi$) is finite, the expression on the right hand side will vanish. To calculate this limit we will use the obtained expression for $P^2$:
\begin{equation}
P^2 = \frac{(1-x^2)\left( (x-\frac{1}{2}\eta n(1-\Lambda'\gamma))^2 +\eta^2 + \frac{1-x^2}{1-\Lambda'\gamma} \right)}{(x-\frac{1}{2}\eta n(1-\Lambda'\gamma))^2+ \eta^2}
\end{equation}
and plug it in the following:
\begin{align}\label{war}
\frac{1}{P^2} dx = \frac{1}{\sin\theta} d\theta.
\end{align}
We can use new parameters: $b:=-\frac{1}{2}\eta n(1-\Lambda'\gamma)$, $g=\frac{1}{1-\Lambda'\gamma}$ and simplify $\frac{1}{P^2}$ as follows:
\begin{align}
\frac{1}{P^2} = \frac{1}{1-x^2} - \frac{g}{x^2(1-g)+2bx+\eta^2+b^2+g}.
\end{align}

That way integrating the left hand side of eq. (\ref{war}) yields:
\begin{align}\label{lhs}
L &= \int \left(\frac{1}{1-x^2} - \frac{g}{x^2(1-g)+2bx+\eta^2+b^2+g}\right)dx \nonumber \\
&= \int \left(\frac{1}{1-x^2} -\frac{g}{(1-g)} \frac{1}{x^2+2x\frac{b}{1-g}+\frac{\eta^2+b^2+g}{1-g}}\right)dx  \nonumber\\
&= \log\left(\sqrt{\frac{x+1}{1-x}}\right) - \frac{2G}{\sqrt{4A-B^2}}\arctan\left( \frac{B+2x}{\sqrt{4A-B^2}} \right) +C \nonumber \\
&= \log\left(C'\sqrt{\frac{x+1}{1-x}}\right) - \frac{2G}{\sqrt{4A-B^2}} \arctan\left( \frac{B+2x}{\sqrt{4A-B^2}} \right),
\end{align}
where:
\begin{align}
G&=\frac{g}{1-g}; \nonumber \\
A&=\frac{\eta^2+b^2+g}{1-g}; \nonumber \\
B&= \frac{2b}{1-g}; \nonumber
\end{align} 
and we assumed that $4A-B^2>0$, otherwise the term under square root would take the form: $-4A+B^2$ and the sign in front of the $\arctan$ function would change. Next we integrate the right hand side of eq. (\ref{war}) to obtain:
\begin{align}\label{rhs}
R = \int \frac{1}{\sin\theta} d\theta = -\log(\cot\theta +\frac{1}{\sin\theta}) + D = \log\left(\frac{\sin\theta}{\cos\theta + 1}\right) + D.
\end{align}
Using expressions (\ref{lhs}) and (\ref{rhs}) we find $\theta$ as a function of $x$:
\begin{align}
\theta (x) = 2 \arctan\left(C'' \sqrt{\frac{x+1}{1-x}} \exp\left( \frac{-2G}{\sqrt{4A-B^2}} \arctan\left(\frac{B+2x}{\sqrt{4A-B^2}} \right)\right) \right).
\end{align}

Next we write $\sin^2\theta$ in terms of $x$:
\begin{align}\label{sin}
\sin^2\theta &= 4\Bigg(C''^2 \frac{x+1}{1-x}\exp\left( \frac{-4G}{\sqrt{4A-B^2}} \arctan\left(\frac{B+2x}{\sqrt{4A-B^2}} \right)\right)  \nonumber \\
&+ \frac{1}{C''^2}\frac{1-x}{x+1}\exp \left( \frac{4G}{\sqrt{4A-B^2}} \arctan\left(\frac{B+2x}{\sqrt{4A-B^2}} \right) \right)+ 2\Bigg)^{-1}.
\end{align}
Finally, we use (\ref{P^2}) and (\ref{sin}) to find:
\begin{align}
\frac{P^2}{\sin^2\theta} &=  \frac{(x+a)^2+\eta^2+g(1-x^2)}{4((x+a)^2+\eta^2)}\Bigg( C''^2 (x+1)^2\exp\left( \frac{-4G}{\sqrt{4A-B^2}} \arctan\left(\frac{B+2x}{\sqrt{4A-B^2}} \right)\right) \nonumber\\
&+ \frac{1}{C''^2}(1-x)^2\exp \left( \frac{4G}{\sqrt{4A-B^2}} \arctan\left(\frac{B+2x}{\sqrt{4A-B^2}} \right) \right)^{-1}+ 2(1-x^2) \Bigg) \nonumber \\
&= \frac{(x+\frac{1}{2}B(1+G)^{-1})^2+(A(1+G)-G(1+G)-\frac{1}{4}B^2)(1+G)^{-2}+G(1+G)^{-1}(1-x^2)}{4((x+\frac{1}{2}B(1+G)^{-1})^2+\eta^2)} \times \nonumber \\ 
&\times \Bigg( C''^2 (x+1)^2\exp\left( \frac{-4G}{\sqrt{4A-B^2}} \arctan\left(\frac{B+2x}{\sqrt{4A-B^2}} \right)\right) \nonumber\\
&+ \frac{1}{C''^2}(1-x)^2\exp \left( \frac{4G}{\sqrt{4A-B^2}} \arctan\left(\frac{B+2x}{\sqrt{4A-B^2}} \right) \right)^{-1}+ 2(1-x^2) \Bigg) \nonumber
\end{align}
therefore the term $\frac{P^2}{\sin\theta}$ is finite at the poles and in the consequence the right hand side of eq. (\ref{domega}) vanishes\footnote{It is easy to see that for $P^2$ of the form (\ref{vanP^2}) the get the same conclusion}.

\end{document}